\documentclass[nofootinbib, reprint, preprintnumbers, amsmath, amssymb, amsfonts, aps, pra, superscriptaddress, floatfix]{revtex4-2}
\usepackage[utf8]{inputenc}
\usepackage{mathtools}
\usepackage{graphicx}
\usepackage{dcolumn}
\usepackage{bm}
\usepackage{physics}
\usepackage{color}
\usepackage{braket}
\usepackage[normalem]{ulem}
\usepackage{enumerate}
\usepackage{enumitem}
\usepackage[colorlinks=true,linkcolor=blue,citecolor=blue,urlcolor =blue]{hyperref}
\usepackage{mathrsfs}
\usepackage{bbm}
\usepackage{siunitx}
\usepackage{ulem}

\usepackage{comment}
\usepackage{lipsum}

\begin{document}
\title{Identification of Phase Plate Properties Using Photonic Quantum Sensor Networks}

\author{Yusuke Machida}
\affiliation{Department of Electrical, Electronic, and Communication Engineering, Faculty of Science and Engineering, Chuo university, 1-13-27, Kasuga, Bunkyo-ku, Tokyo 112-8551, Japan}%

\author{Hiroki Kuji}
\affiliation{Department of Electrical, Electronic, and Communication Engineering, Faculty of Science and Engineering, Chuo University, 1-13-27, Kasuga, Bunkyo-ku, Tokyo 112-8551, Japan}%
\affiliation{Department of Physics, Tokyo University of Science,1-3 Kagurazaka, Shinjuku, Tokyo, 162-8601, Japan}

 \author{Yuichiro Mori}
\thanks{The current affiliation is Department of Electrical, Electronic, and Communication Engineering, Faculty of Science and Engineering, Chuo University, 1-13-27, Kasuga, Bunkyo-ku, Tokyo 112-8551, Japan}
\affiliation{Global Research and Development Center for Business by Quantum-AI Technology (G-QuAT), National Institute of Advanced Industrial Science and Technology (AIST), 1-1-1, Umezono, Tsukuba, Ibaraki 305-8568, Japan}%

 \author{Hideaki Kawaguchi}
\affiliation{Medical Research Center for Pre-Disease State (Mebyo) AI, Graduate School of Medicine, The University of Tokyo, 7-3-1, Hongo, Bunkyo-ku, Tokyo 113-0033, Japan}
\affiliation{Graduate School of Science and Technology, Keio University, Yokohama, Kanagawa 223-8522 Japan}%

 \author{Takashi Imoto}
\affiliation{Global Research and Development Center for Business by Quantum-AI Technology (G-QuAT), National Institute of Advanced Industrial Science and Technology (AIST), 1-1-1, Umezono, Tsukuba, Ibaraki 305-8568, Japan}%

\author{Yuki Takeuchi}
\thanks{The current affiliation is Information Technology R\&D Center, Mitsubishi Electric Corporation.}
\affiliation{NTT Communication Science Laboratories, NTT Corporation,3-1 Morinosato Wakamiya, Atsugi, Kanagawa 243-0198, Japan
}%
\affiliation{NTT Research Center for Theoretical Quantum Information, NTT Corporation,3-1 Morinosato Wakamiya, Atsugi, Kanagawa 243-0198, Japan}

\author{Miku Ishizaki}
\affiliation{Department of Electrical, Electronic, and Communication Engineering, Faculty of Science and Engineering, Chuo university, 1-13-27, Kasuga, Bunkyo-ku, Tokyo 112-8551, Japan}%

\author{Yuichiro Matsuzaki}
\affiliation{Department of Electrical, Electronic, and Communication Engineering, Faculty of Science and Engineering, Chuo university, 1-13-27, Kasuga, Bunkyo-ku, Tokyo 112-8551, Japan}%

\date{\today}


\begin{abstract}

Quantum sensor networks (QSNs) have been widely studied for their potential
of precise measurements. While most QSN research has focused on estimating continuous variables, recent studies have explored discrete-variable estimation. 
Here, we propose a method for high-precision identification of phase plate properties using a photon-based QSN, which is categorized as discrete-variable estimation. 
We 
consider an interaction of a single photon with $N$ phase plates. 
There are some distinct properties of the phase plates, and we aim to identify such properties.
Specifically, we investigate two cases: (i) distinguishing between phase plates that impart uniformly random phases in the range $[0, 2\pi]$ and those that impart the same phase, and (ii) distinguishing between phase plates that impart uniformly random phases in $[0, 2\pi]$ and those that impart phases within a narrower range $[- \delta, \delta]$ ($0< \delta \ll 1$).  
For this distinction, we consider two approaches: one in which a single photon is prepared in a nonlocal state before interacting with the phase plates, and the other in which the single photon remains in a local state.
Our results demonstrate that the nonlocal state enables more precise identification when $N$ is large. 

\end{abstract}

\maketitle
\section{Introduction}
\label{sec:intro}
Quantum sensors have been widely studied as an application of quantum mechanics~\cite{degen2017quantum}. Due to their high sensitivity to external interference, quantum sensors can measure various physical phenomena with extremely high precision. In classical physics, measurement errors cannot exceed the standard quantum limit (SQL). In contrast, the quantum theory allows for surpassing the SQL by leveraging entanglement, and the ultimate precision limit in such cases is known as the Heisenberg limit ~\cite{huelga1997improvement,bollinger1996optimal,wineland1992spin,nagata2007beating,kok2017role,giovannetti2011advances}.  

Quantum sensors can be categorized into those utilizing solid-state qubits and those utilizing 
photons. In solid-state qubit-based sensors, the superposition 
between a ground state and an excited state
acquires a relative phase because of external fields, enabling efficient field-strength estimation through qubit measurements ~\cite{maze2008nanoscale,balasubramanian2008nanoscale}. Such sensors are useful for estimating magnetic fields, electric fields, and temperature ~\cite{barry2020sensitivity}. Furthermore, using solid-state qubits on a nanoscale as sensors is expected to improve spatial resolution ~\cite{schaffry2011proposed,maletinsky2012robust}.  
In addition, quantum sensors can perform optical phase measurements ~\cite{giovannetti2011advances}. 
High-precision optical phase measurement has numerous important applications, including medical and biological sensing ~\cite{taylor2016quantum}. Furthermore, results exceeding the SQL have been demonstrated using two-photon and four-photon states ~\cite{nagata2007beating,polino2020photonic,kok2017role}. In this paper, we focus on the latter type of quantum sensor based on photon.

In recent years, applications of connecting quantum sensors to quantum networks have been actively studied ~\cite{zhan2024optimizing,zang2024quantum,knott2016local,humphreys2013quantum,takeuchi2019quantum,bate2025experimental,qian2019heisenberg,bringewatt2021protocols,shettell2020graph,eldredge2018optimal,okane2021quantum,kasai2024direct,kasai2022anonymous,yin2020experimental}. A quantum sensor network (QSN) consists of four key steps: initialization of the quantum sensor state, interaction between the quantum sensor and the measured system, measurement on the quantum sensor, and classical processing. In certain scenarios within a QSN, using an entangled initial state enables more accurate estimation compared to a non-entangled state ~\cite{zhan2024optimizing,zang2024quantum,knott2016local,humphreys2013quantum,liu2021distributed,rubio2020quantum}. A notable example is the optical atomic clock, where entangling multiple clocks enhances both precision and stability ~\cite{komar2014quantum}.  

Most conventional studies on QSN have focused on estimating phase values, which are continuous variables ~\cite{humphreys2013quantum}. However,
there are some recent studies about
estimating discrete values instead of continuous parameters ~\cite{hillery2023discrete,zhan2024optimizing,ali2024discrete}. For example, consider a setup where multiple detectors are prepared, with a unitary operator acting on only one of them while identity operators act on the others ~\cite{zhan2024optimizing}. By selecting an appropriate initial state and observable, it is possible to determine which detector the unitary operator has acted upon ~\cite{zhan2024optimizing}. In such a setup, the quantity to be estimated is discrete, distinguishing it from traditional protocols that estimate continuous-phase values.

\begin{figure}[h!t]
    \centering
        \includegraphics[width=0.55\textwidth]{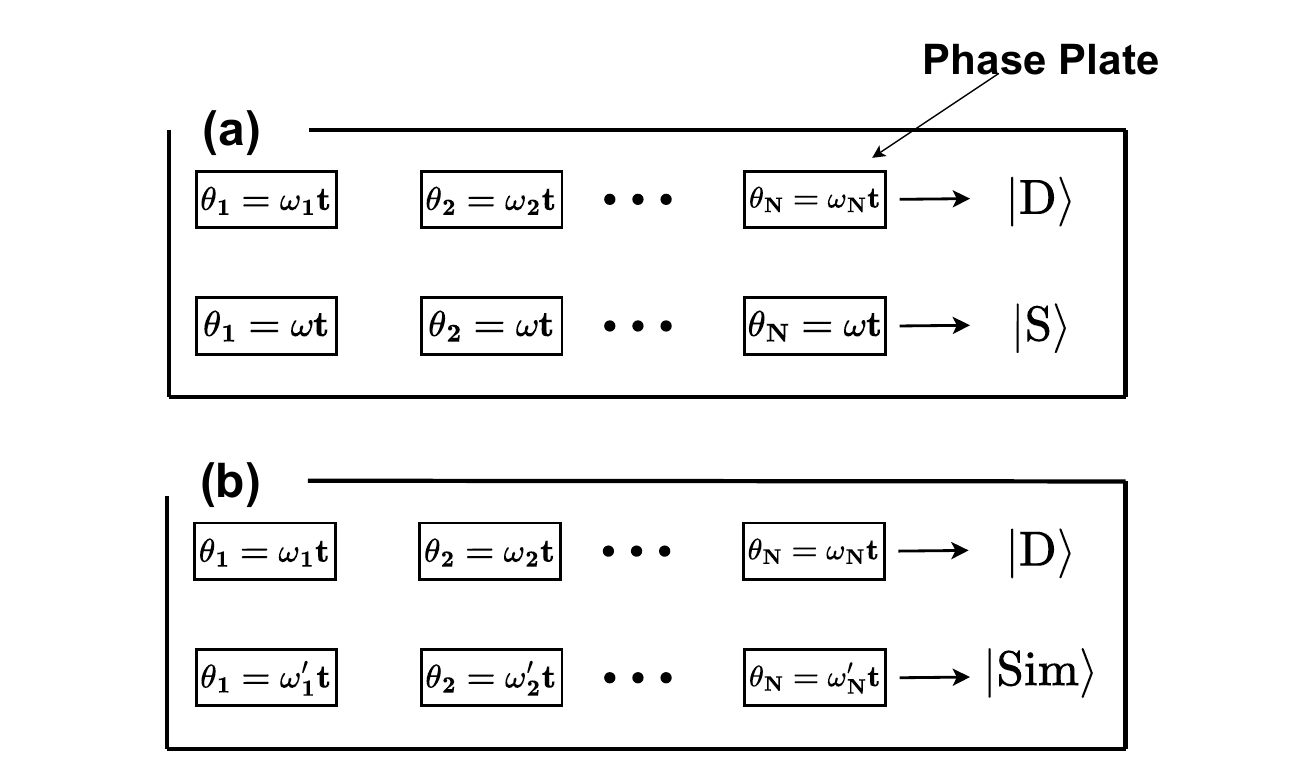}
        \caption{In our setup, there are some distinct properties of the phase plates where we aim to identify the properties. Specifically, we consider two cases.
        In the first case (a), we aim to
        distinguish between phase plates  
        that impart uniformly random phases in the range \([0, 2\pi]\)
        (\(\ket{\rm{D}}\)) 
        and those that impart the same phase (\(\ket{\rm{S}}\)). In the second case (b), we 
        aim to distinguish between
        phase plates 
        that impart uniformly random phases in the range \([0, 2\pi]\)
        (\(\ket{\rm{D}}\)) and those that impart phases which are non-uniform but have only a small degree of variation (\(\ket{\rm{Sim}}\)).  
        We propose a method for identifying which type of phase plates has been prepared in each case (a) and (b) using a single photon.} 
        \label{fig:two_case} 
\end{figure}

Here, we propose a method for high-precision identification of phase plate properties using a quantum sensor network (QSN) based on photons, which is categorized as discrete-variable estimation. Specifically, we 
investigate the behavior of a single photon in a system with \(N\) phase plates. 
There are some distinct properties of the phase plates, and we aim to identify such properties by using a single photon.
We consider two cases: first, distinguishing between phase plates that impart uniformly random phases in the range \([0, 2\pi]\) and those that impart the same phase (Fig.\ref{fig:two_case} (a)); second, distinguishing between phase plates that impart uniformly random phases in the range \([0, 2\pi]\) and those that impart uniformly random phases within the narrower range \([- \delta, \delta]\), where \(\delta\) is a sufficiently small value (\(0<\delta \ll 1\)) (Fig.\ref{fig:two_case} (b)).  
We send a single photon to interact with the phase plates, and measure the photon to distinguish such properties of the phase plates. In particular, we
adopt two strategies:
 one in which a single photon is split by beam splitters and prepared in a nonlocal state before interacting with the phase plates, and another in which a local single-photon state is prepared before interaction. Our results demonstrate that the former approach to use the nonlocal states enables more precise identification when \(N\) is large.  
Our findings introduce a new
application for discrete variable measurements using QSN.  

\begin{figure}[h!t]
    \centering
        \includegraphics[width=0.55\textwidth]{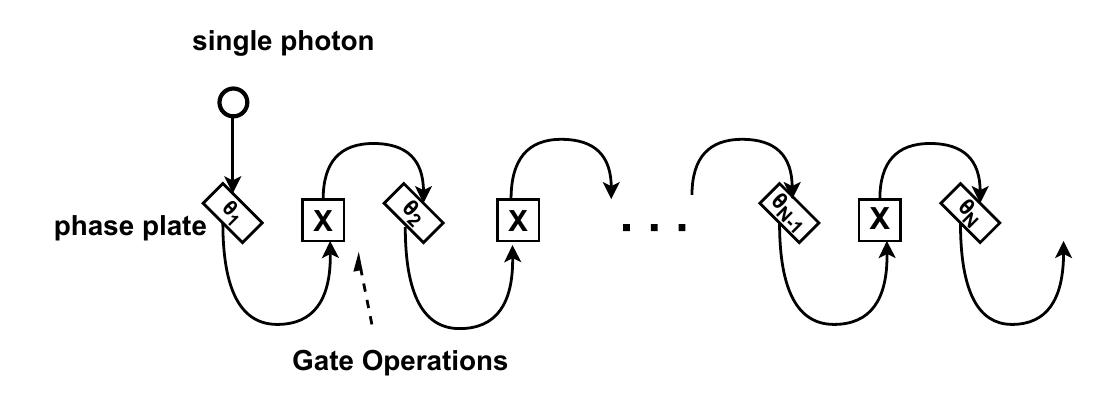}
        \caption{A single photon is initially prepared in a local state and then interacts sequentially with \(N\) phase plates. The photon interacts with one phase plate at a time, and a bit-flip operation is performed after each interaction. This process is repeated sequentially for a total of \(N - 1\) times. After interacting with the final phase plate, a measurement is performed on the state.} 
        \label{fig:kyokusyo} 
\end{figure}

\begin{figure}[h!t]
    \centering
        \includegraphics[width=0.52\textwidth]{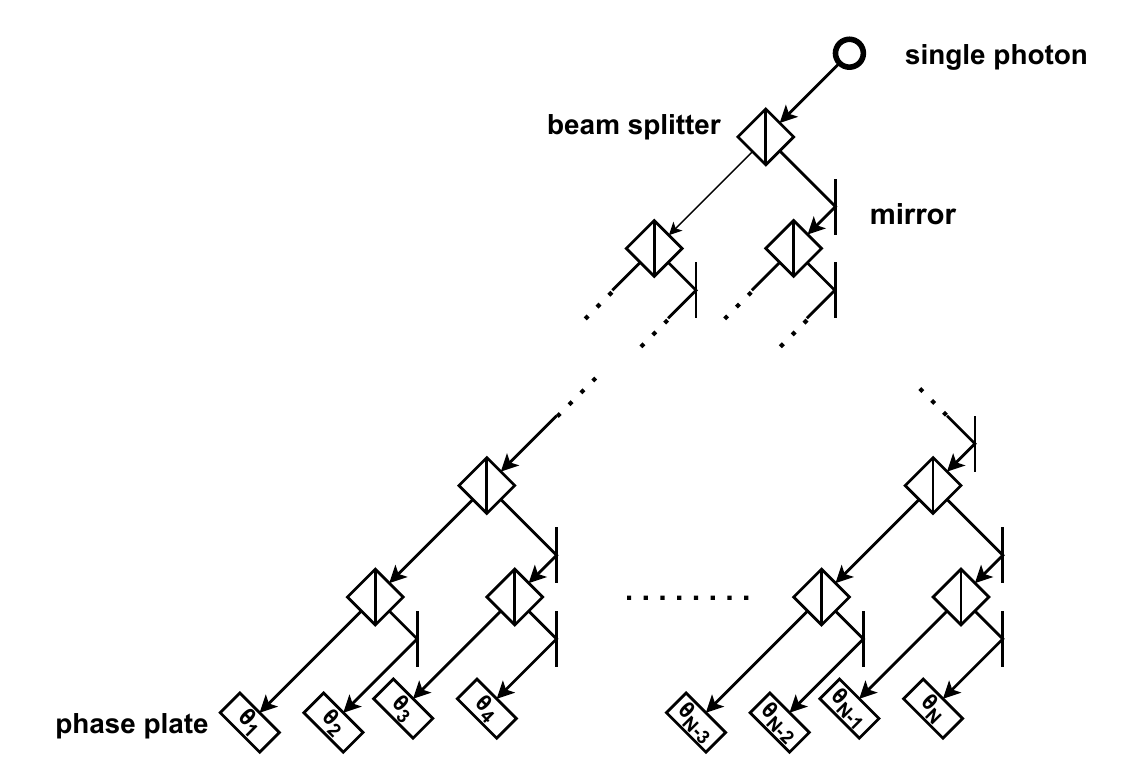}
        \caption{A single photon is initially prepared in a nonlocal state and then interacts with \(N\) phase plates. The photon is first split into \(N\) optical paths using \(N - 1\) beam splitters. Each path then enters a corresponding phase plate and interacts with it. After these interactions, a measurement is performed on the final state.} 
        \label{fig:hikyokusyo} 
\end{figure}


\section{Model and Results}
\subsection{Identification of Phase Plates Imparting Uniform and Non-Uniform Phases} \label{kvsh}
\subsubsection{setup}
We consider a system consisting of \(N\) phase plates and the optical modes that interact with them. For simplicity, we assume that \(N\) is even and adopt the single-rail encoding scheme, using a single photon ~\cite{lee2000quantum,lund2002nondeterministic,drahi2021entangled}. In this encoding, the absence of a photon in a given mode is represented as \(|0\rangle_j\), while the presence of a single photon is represented as \(|1\rangle_j\), defining a qubit. Here, \(j\) denotes the label of the corresponding phase plate to be interacted with the photon, where \(j = 1,2,\cdots, N\). The Pauli \(Z\) operator is then defined as \(\hat{\sigma}^{(j)}_z = |1\rangle_j\langle1| - |0\rangle_j\langle0|\).  

We consider two distinct
cases: one where the phase plates introduce non-uniform phases and the other where all phase plates impart the same phase (Fig.\ref{fig:two_case} (a)). The former case is denoted as \(|\rm{D}\rangle\) (Different), and the latter is denoted as \(|\rm{S}\rangle\) (Same).  
We assume that, at the beginning, it is unknown which setup has been prepared, and thus the initial state is given by the mixed state \(\frac{1}{2} |\rm{D}\rangle \langle \rm{D}| + \frac{1}{2} |\rm{S}\rangle \langle \rm{S}|\). These two setups are treated as orthogonal quantum states such as $\langle \rm{D} |\rm{S}\rangle =0$.  
We aim to determine which setup was prepared by allowing a single photon to interact with the phase plates and then measuring it. In particular, we aim to achieve high-precision discrimination with a single measurement on the photon.

\subsubsection{Interaction of a Local Single-Photon State with \(N\) Phase Plates} \label{kyokusyokekka}
 
First, we analyze the case where a single photon in a local state interacts with phase plates (see Fig.\ref{fig:kyokusyo}). A single photon interacts with a phase plate, and subsequently, a quantum bit-flip operation is applied. This process is repeated multiple times. As described later, in this method, if the phase plate imparts the same phase, the output state of the single photon remains the same as its initial state. On the other hand, if the phase plate imparts different phases, the output state may deviate from the initial state. This property is used to distinguish between the two setups.

The Hamiltonian of the system is defined as:
\begin{align}
    \hat{H}=&\left(\sum_{j=1}^{N}\frac{\hat{1}_j+\hat{\sigma}_{z}^{(j)}}{2}\omega_{j}\right)\otimes\ket{\rm{D}}\bra{\rm{D}}  \notag \\
    &+\left(\sum_{j=1}^{N}\frac{\hat{1}_j+\hat{\sigma}_z^{(j)}}{2}\omega\right)\otimes\ket{\rm{S}}\bra{\rm{S}},
    \label{hamiltoniam}
\end{align}
where  $\omega_j$ represents the phase acquired by the single photon per unit time as it passes through the
$j$-th phase plate,
$\hat{1}_{j}$
($j=1,2,\cdots ,N$) is the identity operator corresponding to the $j$-th system mode, and $\hat{1}_{N+1}=\ket{\rm{D}}\bra{\rm{D}}+\ket{\rm{S}}\bra{\rm{S}}$.
Let
$t$  denote the interaction time with the phase plate. In the first term on the right-hand side,  
$\theta_j=\omega_jt$
represents the phase acquired by the single photon from the  
$j$-th phase plate, which is assumed to be inhomogeneous. In the second term, the phase acquired by the single photon is uniform and described as  
$\theta=\omega t$ .

The initial state $\rho_0^{(\rm{local})}$is given by:

\begin{align}
    \rho_0^{(\rm{local})}=\ket{+}_1\bra{+}\otimes\ket{\Tilde{0}}_{2,3\cdots N}\bra{\Tilde{0}}\otimes\left(\frac{1}{2}\ket{\rm{D}}\bra{\rm{D}}+\frac{1}{2}\ket{\rm{S}}\bra{\rm{S}}\right),
\end{align}
where $|+\rangle =\frac{1}{\sqrt{2}}(|0\rangle +|1\rangle)$ and $\ket{\Tilde{0}}=\ket{0}_2\ket{0}_3\cdots\ket{0}_N$.
Here, the single photon acts as a flying qubit that moves through space as time progresses. After interacting with the phase plate fixed at a specific position, the single photon moves to the next plate, and SWAP operations represent this movement.

After repeated application of the time evolution (dictated by the given Hamiltonian), quantum bit-flip operations, and SWAP operations, the state $\rho_t^{(\rm{local})}$at time $t$ becomes:
\begin{align}
    \rho_t^{(\rm{local})}=\hat{V}\rho_0\hat{V}^{\dagger},
\end{align}
where $\hat{V}$ and $\hat{V}_j$ are defined as follows:
\begin{align}
    \hat{V}&=\exp[-i\hat{H}t]\hat{V}_{N-1}\cdots\hat{V}_{2}\hat{V}_{1},\\
    \hat{V}_j&=\hat{U}_{SWAP}^{(j,j+1)}\hat{\sigma}_x^{(j)}\exp[-i\hat{H}t].
\end{align}
where $\hat{U}_{SWAP}^{(j,j+1)}$ denotes a SWAP gate between $j$-th qubit and $(j+1)$-th qubit.
So
We obtain the following
\begin{align}
    \notag\rho_t^{(\rm{local})} 
    \notag
    &=\frac{1}{2}\ket{\Tilde{0}^{\prime}}_{1,2\cdots N-1}\bra{\Tilde{0}^{\prime}}\otimes\ket{+}_N\bra{+}\otimes\ket{\rm{S}}\bra{\rm{S}} \\
    &+\frac{1}{2}\ket{\Tilde{0}^{\prime}}_{1,2\cdots N-1}\bra{\Tilde{0}^{\prime}}\otimes\ket{\psi}_N\bra{\psi}\otimes\ket{\rm{D}}\bra{\rm{D}},
    \label{eq:DenseMatrix_system}
\end{align}
where $\ket{\Tilde{0}^{\prime}}=\ket{0}_1\ket{0}_2\cdots\ket{0}_{N-1}$, and  $\ket{\psi}$ is expressed as:
\begin{align}
    \notag \ket{\psi}=&\frac{1}{\sqrt{2}} \exp[-it(\omega_{1}+\omega_{3}+\cdots+\omega_{N-1})]\ket{0} \\ 
    &+\frac{1}{\sqrt{2}}\exp[-it(\omega_{2}+\omega_{4}+\cdots+\omega_{N})]\ket{1} .
\end{align}

Let $\rho_{\rm{S}}^{(\rm{local})}$ and $\rho_{\rm{D}}^{(\rm{local})}$ denote the states that evolve when the phase plates correspond to $\ket{\rm{S}}$ and $\ket{\rm{D}}$, respectively. These are described as:

\begin{align}
    \rho_{\rm{S}}^{(\rm{local})}&=\ket{\Tilde{0}^{\prime}}_{1,2,\cdots N-1}\bra{\Tilde{0}^{\prime}}\otimes \ket{+}_{N}\bra{+}\otimes\ket{\rm{S}}\bra{\rm{S}},\\
    \rho_{\rm{D}}^{(\rm{local})}&=\ket{\Tilde{0}^{\prime}}_{1,2,\cdots N-1}\bra{\Tilde{0}^{\prime}}\otimes \ket{\psi}_{N}\bra{\psi}\otimes\ket{\rm{D}}\bra{\rm{D}}.
\end{align}
We define projection operators as follows:
\begin{align}
    \hat{P}_{\rm{S}}^{(\rm{local})}&=\hat{1}_{1,2,\cdots N-1} \otimes \ket{+}_N\bra{+} \otimes \hat{1}_{N+1}, \\
    \hat{P}_{\rm{D}}^{(\rm{local})}&=\hat{1}_{1,2,\cdots N-1} \otimes \ket{-}_N\bra{-} \otimes \hat{1}_{N+1},
\end{align}
where $\hat{1}_{1,2,\cdots,N-1}=\hat{1}_1\otimes \hat{1}_2 \otimes \cdots \otimes \hat{1}_{N-1}$.
If the system is projected 
by $\hat{P}_{\rm{S}}$, the phase plate is identified as providing a uniform phase. Conversely, if it is projected 
by $\hat{P}_{\rm{D}}$, the phase plate is identified as providing an 
inhomogeneous phase. The reason for this definition is that, in the case of a uniform phase plate, the probability of projection onto $\hat{P}_{\rm{D}}$ is 0, whereas, for an 
inhomogeneous phase plate, projection by $\hat{P}_{\rm{D}}$ occurs with a finite probability (as discussed later).

In conventional quantum sensor networks, quantum fisher information has been used to quantify sensitivity for continuous measurements~\cite{humphreys2013quantum}. However, since our method handles discrete variables, quantum fisher information cannot be directly applied. Instead, we employ the error probability to evaluate the performance of distinguishing between two different states ~\cite{barnett2017introduction,helstrom1969quantum,helstrom1967detection}.
Such an error probability $P_{\rm{err}}$ is defined as:
\begin{align}
    P_{\rm{corr}}=&\sum_{j={\rm{S}},{\rm{D}}} p_j \operatorname{Tr}[\hat{P}_j \rho_j] , \\
     P_{\rm{err}}=&1-P_{\rm{corr}}
    =1-\sum_{j={\rm{S}},{\rm{D}}} p_j \operatorname{Tr}[\hat{P}_j \rho_j] ,\label{eq:ayatei}
\end{align}
where $p_{\rm{S}}$ and $p_{\rm{D}}$ represent prior probabilities. Since it is initially unknown whether the phase plate imparts uniform or 
inhomogeneous phases, we assume prior probabilities of $p_{\rm{S}}=p_{\rm{D}}=1/2$.
In prior research ~\cite{helstrom1969quantum,helstrom1967detection}, error probability was used to differentiate between the presence or absence of interaction with a target signal. In contrast, our method employs error probability to quantify the accuracy of distinguishing between states resulting from interaction with uniform versus inhomogeneous phase plates.

We calculate the value of the error probability when the photon is in a local state.

\begin{align}
    P_{\rm{err}}^{(\rm{local})}=\frac{1}{2}\left|\braket{+|\psi}\right|^2. \label{isiayakyoku}
\end{align}
To do this, we evaluate $\left|\braket{+|\psi}\right|^2$, which is given by:
\begin{align}
    \left|\braket{+|\psi}\right|^2=\cos^2
    \Big{(}
    \frac{\omega_1-\omega_2+\omega_3-\omega_4+\cdots+\omega_{N-1}-\omega_{N}}{2}t\Big{)}.
\end{align}
Also, we obtain
\begin{align}
    \left|\braket{-|\psi}\right|^2=\sin^2
    \Big{(}
    \frac{\omega_1-\omega_2+\omega_3-\omega_4+\cdots+\omega_{N-1}-\omega_{N}}{2}t\Big{)}.
\end{align}
This means that
when the phase plate imparts
inhomogeneous phases, projection by $\hat{P}_{\rm{D}}$ occurs with a finite probability. 
Assuming $\theta_1,\theta_2,\cdots,\theta_N$ are independently and uniformly drawn from the interval  $[0,2\pi]$ , the statistical average of $ P_{\rm{err}}^{(\rm{local})}$is given as:
\begin{align}
    &\langle P_{\rm{err}}^{(\rm{local})}\rangle \notag\\
    =&\frac{1}{2}\frac{1}{(2\pi)^N}\int^{2\pi}_{0}\int^{2\pi}_{0}\cdots\int^{2\pi}_{0}\notag \\
    &\cos^2\Big{(}\frac{\theta_1-\theta_2+\theta_3-\theta_4+\cdots+\theta_{N-1}-\theta_{N}}{2}\Big{)}\notag d\theta_1d\theta_2\cdots d\theta_N\\
    =&\frac{1}{4}. \label{eq:isiayakyoanakVSh}
\end{align}
This result indicates that the average of the error probability cannot be reduced below $1/4$ in the case of the local states.

\subsubsection{Interaction of a nonlocal Single-Photon State with \(N\) Phase Plates}\label{hikyokusyokekka}

Next, we consider the case where a single photon in a nonlocal state is used (see Fig.\ref{fig:hikyokusyo}). Specifically, the same Hamiltonian described in \ref{kyokusyokekka} (Equation \eqref{hamiltoniam}) is used, but the initial state $\rho_0^{(\rm{nonlocal})}$ is defined as follows:
\begin{align}
&\rho_0^{(\rm{nonlocal})}=\ket{\psi_0}\bra{\psi_0}\otimes\left(\frac{1}{2}\ket{\rm{D}}\bra{\rm{D}}+\frac{1}{2}\ket{\rm{S}}\bra{\rm{S}}\right),\\
    &\ket{\psi_0}=\frac{1}{\sqrt{N}}\left(\ket{0\cdots01}+\ket{0\cdots10}+\cdots+\ket{1\cdots00} \right),
\end{align}
where $\ket{\psi_0}$ is a W-state involving $N$ modes. The W-state for $N$ modes can be generated by injecting a single photon into a series of beam splitters. For example, a single photon injected into one beam splitter generates a W-state spanning two modes. Injecting it into two beam splitters generates a W-state spanning four modes. By repeating this process $m$ times, an $N=2^m$ mode W-state can be produced ($m$ being a natural number). While prior research has explored using W-states as probes in QSN, those studies focused on measuring continuous variables such as phase ~\cite{maleki2022distributed}. In contrast, our approach leverages W-states to distinguish between two cases, which is a key difference.

After time evolution under the given Hamiltonian for a duration 
$t$, the resulting state $\rho_t$ is expressed as
\begin{align}
    \rho_t^{(\rm{nonlocal})}=e^{-i\hat{H}t}\rho_0e^{i\hat{H}t}.
\end{align}
Let us define $\ket{w}$ and $\ket{w_{\rm{D}}}$ as follows:
\begin{align}
    &\ket{w} = \frac{1}{\sqrt{N}}(\ket{0\cdots01}+\ket{0\cdots10}+\cdots+\ket{1\cdots00})=\ket{\psi_0},\\
    \notag&\ket{w_{\rm{D}}}
   =\frac{1}{\sqrt{N}}(\exp[-it\omega_N]\ket{0\cdots01}\\&+\exp[-it\omega_{N-1}]\ket{0\cdots10}  
    +\cdots+\exp[-it\omega_1]\ket{1\cdots00}).
\end{align}
By expanding $\rho_t$, it can be written as:
\begin{align}
     \rho_t^{(\rm{nonlocal})}=\frac{1}{2}\ket{w}\bra{w}\otimes\ket{\rm{S}}\bra{\rm{S}}+\frac{1}{2}\ket{w_{\rm{D}}}\bra{w_{\rm{D}}}\otimes\ket{\rm{D}}\bra{\rm{D}}.
\end{align}
Also, we calculate the error probability when
we aim to identify the properties of the phase plates using a single photon in a nonlocal state. Let $\rho_{\rm{S}}^{(\rm{nonlocal})}$ denote the state when the plate provides $\ket{{\rm{S}}}$, and let $\rho_{\rm{D}}^{(\rm{nonlocal})}$ denote the state when the plate provides $\ket{{\rm{D}}}$.
These states are expressed as follows:
\begin{align}
    \rho_{\rm{S}}^{(\rm{nonlocal})}&=\ket{w}\bra{w}\otimes\ket{{\rm{S}}}\bra{{\rm{S}}},\\
    \rho_{\rm{D}}^{(\rm{nonlocal})}&=\ket{w_{\rm{D}}}\bra{w_{\rm{D}}}\otimes\ket{{\rm{D}}}\bra{{\rm{D}}}.
\end{align}

The corresponding projection operators are defined as follows:

\begin{align}
    \hat{P}_{\rm{S}}^{(\rm{nonlocal})}&=\ket{w}\bra{w} \otimes \hat{1}_{N+1}, \\
    \hat{P}_{\rm{D}}^{(\rm{nonlocal})}&= (\hat{1}_{1,2,\cdots,N}-\ket{w}\bra{w}) \otimes \hat{1}_{N+1}.
\end{align}
We consider that,
if the system is projected onto $\hat{P}_{\rm{S}}^{(\rm{nonlocal})}$, the plate is identified as providing a uniform phase. Conversely, we consider that,
if the system is projected by $\hat{P}_{\rm{D}}^{(\rm{nonlocal})}$, the plate is identified as providing a 
inhomogeneous phase. This definition is justified because, in the case of a uniform phase plate, the probability of projection by $\hat{P}_{\rm{D}}^{(\rm{nonlocal})}$ is 0, whereas, for a 
inhomogeneous
phase plate, the probability of projection onto $\hat{P}_{\rm{D}}^{(\rm{nonlocal})}$ is 
high (as explained later). The error probability $P_{\rm{err}}$ is adopted in the same form as in the local state case, as given by Equation \eqref{eq:ayatei}.
Similarly, the prior probabilities are assumed to be $p_{\rm{S}}=p_{\rm{D}}=1/2$.
To calculate $P_{\rm{err}}^{(\rm{nonlocal})}$:

\begin{align}
    P_{\rm{err}}^{(\rm{nonlocal})}=\frac{1}{2}\left|\braket{w|w_{\rm{D}}}\right|^2.  \label{eq:isiayahikyoku}
\end{align}
we first determine the value of $\left|\braket{w|w_{\rm{D}}}\right|^2$:

\begin{align}
    \left|\braket{w|w_{\rm{D}}}\right|^2=\left|\frac{1}{N}\sum_{j=1}^{N}\exp[it\omega_j]\right|^2. \label{eq:ww'}
\end{align}
Assuming that $\theta_1,\theta_2,\cdots,\theta_N$ are independently and uniformly distributed over$[0,2\pi]$ , we compute the statistical average of $\left|\braket{w|w_{\rm{D}}}\right|^2$  using Equation \eqref{eq:ww'}:
\begin{align}
    \notag&\left\langle \left| \frac{1}{N}\sum_{j=1}^N \exp[i\hat{\theta}_j] \right|^2 \right\rangle 
    \notag=\left\langle \frac{1}{N^2}\sum_{j,j^{\prime}=1}^N \exp[i\hat{\theta}_j]\exp[-i\hat{\theta}_{j^{\prime}}] \right\rangle \\
    &=\left\langle \frac{1}{N^2}\left( \sum_{j=j^{\prime},j=1}^N 1 + \sum_{j\neq j^{\prime}}  \exp[i\hat{\theta}_j]\exp[-i\hat{\theta}_{j^{\prime}}] \right)\right\rangle=\frac{1}{N},
\end{align}
where $\hat{\theta}_j$ represents a classical random variable, and it is assumed that the phases chosen for different plates are independent. Given that $\theta_j$ are uniformly distributed between $0$ and $2\pi$, the statistical average of $\exp[i\hat{\theta}_j]\exp[-i\hat{\theta}_{j^{\prime}}]$becomes 0.
Using this relationship, we find:
\begin{align}
   \left\langle\exp[i\hat{\theta}_j]\exp[-i\hat{\theta}_{j^{\prime}}]\right\rangle&=\left\langle\exp[i\hat{\theta}_j]\right\rangle 
    \left\langle\exp[-i\hat{\theta}_{j^{\prime}}]\right\rangle=0.
\end{align}
From the above, the statistical average of the error probability is calculated as:
\begin{align}
    \langle P_{\rm{err}}^{(\rm{nonlocal})} \rangle=\frac{1}{2N}. \label{eq:isiayahikyokukai}
\end{align}
Therefore, in the limit of sufficiently large $N$ , the value of $P_{\rm{err}}^{(\rm{nonlocal})}$ asymptotically approaches 0. This indicates that increasing the number of phase plates makes it possible to identify the properties of the phase plates almost without error. 

\subsubsection{Comparison}  \label{hikaku}

\begin{figure}[h!]
    \centering
    \includegraphics[width=0.48\textwidth]{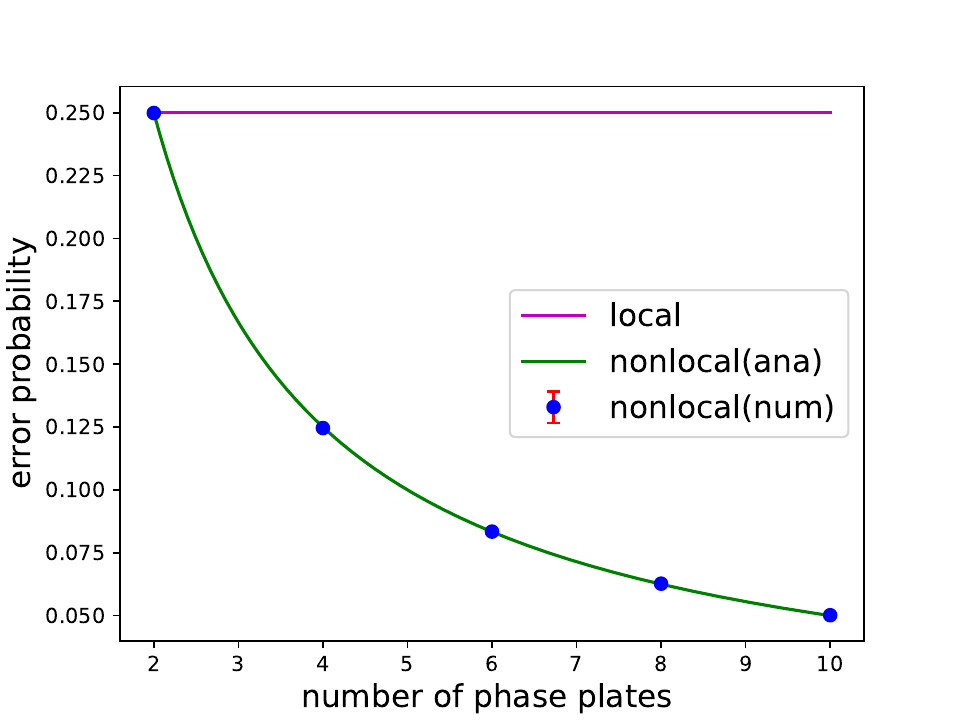}
    \caption{The plot shows the error probabilities for both the local and nonlocal states. The number of phase plates considered ranges from 2 to 10, with \( t = 1 \).  
    For the nonlocal state, the error probability is obtained by numerically computing the average of equation \eqref{eq:isiayahikyoku} over 100,000 trials, and the results are plotted.  
    For the nonlocal state, the analytical solution given by equation \eqref{eq:isiayahikyokukai} is plotted.
    For the local state, the analytical solution given by equation \eqref{eq:isiayakyoanakVSh} is plotted.  
    The standard error is included as error bars in the plot. However, the values are so small that the markers almost hide them.}
    \label{fig:FP_l}
\end{figure}
We compare the error probabilities for 
a local state with that for a nonlocal state (Fig.\ref{fig:two_case}(a)) by using numerical simulations. 
Here, the 
inhomogeneous phases \( \hat{\theta}_j \) are randomly selected uniformly from the range \( [0, 2\pi] \) for each iteration, and the error probability is calculated repeatedly. The statistical average over these iterations is then
taken. 

In Fig.\ref{fig:FP_l}, the number of phase plates ranges from 2 to 10. The plotted results show the statistical averages of the error probabilities for the nonlocal state of numerical results (Equation \eqref{eq:isiayahikyoku}), the nonlocal state of analytical solution (Equation \eqref{eq:isiayahikyokukai}), and the local state (Equation \eqref{eq:isiayakyoanakVSh}). When there are two phase plates, no significant difference in the error probabilities is observed. However, as the number of phase plates increases, the error probability for the nonlocal state decreases and converges to 0. 
Therefore, we conclude that the non-local state provides a better identification than the local state for our purpose.

\subsection{Identification of Phase Plates Imparting Different Non-Uniform Phases}

\subsubsection{setup}
Furthermore,
we consider the task of identifying phase plates in a single measurement, distinguishing between two cases: the phase plates impart uniformly random phases over the range $[0,2\pi]$, and the phases imparted by the phase plates are non-uniform but with a sufficiently small range of variation, smaller than 1 (Fig.\ref{fig:two_case}(b)). We denote the state of the phase plates in the first case as $\ket{\rm{D}}$ (Different), and in the second case as $|\rm{Sim}\rangle $ (Similar).
Since it is initially unknown which configuration is being provided for the experiment, the initial state is represented as a statistical mixture: $\frac{1}{2}|\rm{D}\rangle \langle \rm{D}|+ \frac{1}{2}|\rm{Sim}\rangle \langle \rm{Sim}|$.
These two setups are treated as orthogonal quantum states.

\subsubsection{Interaction of a local Single-Photon State with \(N\) Phase Plates}\label{kyokusyokekkahh}

First, we analyze the case where a single photon in a local state interacts with phase plates (see Fig \ref{fig:kyokusyo}). A single photon interacts with one phase plate and then undergoes a quantum bit-flip operation. This process is repeated multiple times. 

The Hamiltonian of the system is defined as follows:

\begin{align}
     \hat{H}^{\prime}=&\left(\sum_{j=1}^{N}\frac{\hat{1}_j+\hat{\sigma}_{z}^{(j)}}{2}\omega_{j}\right)\otimes\ket{\rm{D}}\bra{\rm{D}} \notag \\
    &+\left(\sum_{j=1}^{N}\frac{\hat{1}_{j}+\hat{\sigma}_z^{(j)}}{2}\omega_{j}^{\prime}\right)\otimes\ket{\rm{Sim}}\bra{\rm{Sim}},
    \label{hamiltoniam_HvsH}
\end{align}
where $\omega_j$ is the phase acquired by the single photon per unit time as it passes through the $j$-th phase plate, $\hat{1}_{j}$ ($j=1,2,\cdots, N$) is the identity operator corresponding to the  
$j$-th system mode, and $\hat{1}_{N+1}=\ket{\rm{D}}\bra{\rm{D}}+\ket{\rm{Sim}}\bra{\rm{Sim}}$ is the identity operator for the phase plates.
Let $t$ denote the interaction time with the phase plate, and in the first term on the right-hand side, $\theta_j=\omega_jt$ represents the phase acquired by the single photon from the phase plate. This phase, $\theta_j$, is assumed to be inhomogeneous. In the second term on the right-hand side, the phase acquired by the single photon from the $j$-th phase plate is given by $\theta_{j}^{\prime}=\omega_{j}^{\prime} t$, which is also 
inhomogeneous but assumed to be small, as we will describe.
The initial state $\Tilde{\rho}_0^{(\rm{local})}$ is defined as:
\begin{align}
    \notag \Tilde{\rho}_0^{(\rm{local})}=&\ket{+}_1\bra{+}\otimes\ket{\Tilde{0}}_{2,3\cdots N}\bra{\Tilde{0}} \\
    &\otimes\left(\frac{1}{2}\ket{\rm{D}}\bra{\rm{D}}+\frac{1}{2}\ket{\rm{Sim}}\bra{\rm{Sim}}\right),
\end{align}
where $|+\rangle =\frac{1}{\sqrt{2}}(|0\rangle +|1\rangle)$
and $\ket{\Tilde{0}}=\ket{0}_2\ket{0}_3\cdots\ket{0}_N$.
After repeatedly applying the time evolution dictated by the given Hamiltonian, quantum bit-flip operations, and SWAP operations, the 
state of the system $\Tilde{\rho}_t^{(\rm{local})}$ becomes:
\begin{align}
    \Tilde{\rho}_t^{(\rm{local})}=\hat{V}^{\prime}\Tilde{\rho}_0\hat{V}^{\prime\dagger},
\end{align}
where $\hat{V}^{\prime}$ and $\hat{V}_j^{\prime}$ are defined as follows:
\begin{align}
    \hat{V}^{\prime}&=\exp[-i\hat{H}^{\prime}t]\hat{V}_{N-1}^{\prime}\cdots\hat{V}_{2}^{\prime}\hat{V}_{1}^{\prime},\\
    \hat{V}_j^{\prime}&=\hat{U}_{SWAP}^{(j,j+1)}\hat{\sigma}_x^{(j)}\exp[-i\hat{H}^{\prime}t].
\end{align}
We obtain
\begin{align}
    \notag\tilde{\rho}_t^{(\rm{local})} 
    &=\frac{1}{2}\ket{\Tilde{0}^{\prime}}_{1,2\cdots N-1}\bra{\Tilde{0}^{\prime}}\otimes\ket{\psi^{\prime}}_N\bra{\psi^{\prime}}\otimes\ket{\rm{Sim}}\bra{\rm{Sim}} \\
    &+\frac{1}{2}\ket{\Tilde{0}^{\prime}}_{1,2\cdots N-1}\bra{\Tilde{0}^{\prime}}\otimes\ket{\psi}_N\bra{\psi}\otimes\ket{\rm{D}}\bra{\rm{D}},
    \label{eq:DenseMatrix_system}
\end{align}
where $\ket{\Tilde{0}^{\prime}}=\ket{0}_1\ket{0}_2\cdots\ket{0}_{N-1}$ and the states $\ket{\psi}$ and $\ket{\psi^{\prime}}$ are defined as :

\begin{align}
    \notag \ket{\psi}=&\frac{1}{\sqrt{2}} \exp[-it(\omega_{1}+\omega_{3}+\cdots+\omega_{N-1})]\ket{0} \\ 
    &+\frac{1}{\sqrt{2}} \exp[-it(\omega_{2}+\omega_{4}+\cdots+\omega_{N})]\ket{1} , \\
    \notag \ket{\psi^{\prime}}=&\frac{1}{\sqrt{2}} \exp[-it(\omega_{1}^{\prime}+\omega_{3}^{\prime}+\cdots+\omega_{N-1}^{\prime})]\ket{0} \\ 
    &+\frac{1}{\sqrt{2}}\ \exp[-it(\omega_{2}^{\prime}+\omega_{4}^{\prime}+\cdots+\omega_{N}^{\prime})]\ket{1} .
\end{align}
Let us define $\Tilde{\rho}_{{\rm{Sim}}}^{(\rm{local})}$
($\tilde{\rho}_{\rm{D}}^{(\rm{local})}$)
as the state after the time evolution when the phase plate is prepared as $\ket{\rm{Sim}}$ ($\ket{\rm{D}}$).
These are described as follows:
\begin{align}
    \tilde{\rho}_{{\rm{Sim}}}^{(\rm{local})}&=\ket{\Tilde{0}^{\prime}}_{1,2,\cdots N-1}\bra{\Tilde{0}^{\prime}}\otimes \ket{\psi^{\prime}}_N\bra{\psi^{\prime}}\otimes\ket{\rm{Sim}}\bra{\rm{Sim}},\\
    \tilde{\rho}_{\rm{D}}^{(\rm{local})}&=\ket{\Tilde{0}^{\prime}}_{1,2,\cdots N-1}\bra{\Tilde{0}^{\prime}}\otimes \ket{\psi}_{N}\bra{\psi}\otimes\ket{\rm{D}}\bra{\rm{D}}.
\end{align}

The corresponding projection operators are defined as follows:

\begin{align}
    \hat{\tilde{P}}_{{\rm{Sim}}}^{(\rm{local})}&=\hat{1}_{1,2,\cdots N-1} \otimes \ket{+}_N\bra{+} \otimes \hat{1}_{N+1}, \\
    \hat{\tilde{P}}_{\rm{D}}^{(\rm{local})}&=\hat{1}_{1,2,\cdots N-1} \otimes \ket{-}_N\bra{-} \otimes \hat{1}_{N+1}.
\end{align}
We consider that,
if the system is projected by $\hat{\tilde{P}}_{{\rm{Sim}}}^{(\rm{local})}$, the phase plate is identified as $|\rm{Sim}\rangle $ . Conversely, we consider that,
if the system is projected by $\hat{\tilde{P}}_{\rm{D}}^{(\rm{local})}$ , the phase plate is identified as $|{\rm{D}}\rangle$. The error probability is evaluated using a 
definition similar to that in Section \ref{kvsh}(Eqation \eqref{eq:ayatei}):
\begin{align}
    \tilde{P}_{\rm{corr}}=&\sum_{j={\rm{Sim}},{\rm{D}}} p_j \rm{Tr}[\hat{\tilde{P}}_j \tilde{\rho}_j],\\
    \tilde{P}_{\rm{err}}=&1-\tilde{P}_{\rm{corr}}
    =1-\sum_{j={\rm{Sim}},{\rm{D}}} p_j \rm{Tr}[\hat{\tilde{P}}_j \tilde{\rho}_j].  \label{eq:ayateiFD}
\end{align}
Here, $p_{{\rm{Sim}}}$ and $p_{\rm{D}}$ are prior probabilities.
Since it is initially unknown which setup is present, we assume that $p_{{\rm{Sim}}}=p_{\rm{D}}=1/2$. The calculated values of the error probability are presented below.
\begin{align}
    \tilde{P}_{\rm{err}}^{(\rm{local})}=\frac{1}{2}(1-\left|\braket{+|\psi^{\prime}}\right|^2 +\left|\braket{+|\psi}\right|^2). \label{eq:hVShisiayakyoku}
\end{align}
Also, we calculate the values of $\left|\braket{+|\psi}\right|^2$ and $\left|\braket{+|\psi^{\prime}}\right|^2$.
\begin{align}
    \left|\braket{+|\psi}\right|^2=\cos^2
    \Big{(}
    \frac{\omega_1-\omega_2+\omega_3-\omega_4+\cdots+\omega_{N-1}-\omega_{N}}{2}t\Big{)},\\
    \left|\braket{+|\psi^{\prime}}\right|^2=\cos^2
    \Big{(}
    \frac{\omega_1^{\prime}-\omega_2^{\prime}+\omega_3^{\prime}-\omega_4^{\prime}+\cdots+\omega_{N-1}^{\prime}-\omega_{N}^{\prime}}{2}t\Big{)}.
\end{align}
Under the assumption that $\theta_1,\theta_2\cdots\theta_N$ are uniformly distributed between 0 and  $2\pi$, the statistical average of $ \left|\braket{+|\psi}\right|^2$  is given by:
\begin{align}
    &\langle \left|\braket{+|\psi}\right|^2\rangle\\
    =&\frac{1}{(2\pi)^N}\int^{2\pi}_{0}\int^{2\pi}_{0}\cdots\int^{2\pi}_{0}\notag \\
    &\cos^2\Big{(}\frac{\theta_1-\theta_2+\theta_3-\theta_4+\cdots+\theta_{N-1}-\theta_{N}}{2}\Big{)}\notag d\theta_1d\theta_2\cdots d\theta_N\\
    =&\frac{1}{2}. \label{eq:isiayakyoana}
\end{align}
Similarly, the calculation of $\left|\braket{+|\psi^{\prime}}\right|^2$ is given as follows: 
\begin{align}
    \left|\braket{+|\psi^{\prime}}\right|^2&=\frac{1}{2}(1+\cos\phi)\\
    &\simeq
    1-\frac{1}{4}\phi^2,
\end{align}
where $\phi$ is defined as:
\begin{align}
    \phi=(\theta_1^{\prime}-\theta_2^{\prime})+(\theta_3^{\prime}-\theta_4^{\prime})+\cdots+(\theta_{N-1}^{\prime}-\theta_N^{\prime}),
\end{align}
and the statistical average of $\phi^2$ is given as:
\begin{align}
    \langle \phi^2 \rangle =\frac{N}{3}\frac{\pi^2}{M^2}.
\end{align}
Here, $\theta_1^{\prime},\theta_2^{\prime}\cdots$ and $\theta_N^{\prime}$ are assumed to be uniformly distributed within $[-\frac{\pi}{M},\frac{\pi}{M}]$, and $M$ is assumed to be 
much larger than 1. The statistical average of $\left|\braket{+|\psi^{\prime}}\right|^2$ is then calculated as follows:
\begin{align}
    \langle\left|\braket{+|\psi^{\prime}}\right|^2\rangle\simeq
    1-\frac{N}{12}\frac{\pi^2}{M^2}.
\end{align}

Using these results, the statistical average of $\tilde{P}_{\rm{err}}^{(\rm{local})}$ is obtained as:

\begin{align}
    \langle  \tilde{P}_{\rm{err}}^{(\rm{local})} \rangle\simeq\frac{N}{24}\frac{\pi^2}{M^2}+\frac{1}{4}. \label{eq:hVShisiayakyokuana}
\end{align}

Consequently, it can be concluded that in the case of the local states, the average of the error probability cannot be
smaller than $1/4$.
\subsubsection{Interaction of a nonlocal Single-Photon State with \(N\) Phase Plates} \label{hikyokusyokekkahh}
We consider the case of using single photons in a nonlocal state (see Fig \ref{fig:hikyokusyo}). The same Hamiltonian as in the local state case (Equation \eqref{hamiltoniam_HvsH}) is employed. The initial state $\tilde{\rho}_0^{(\rm{nonlocal})}$ is defined as follows:
\begin{align}
    \tilde{\rho}_0^{(\rm{nonlocal})}&=\ket{w}\bra{w}\otimes\left(\frac{1}{2}\ket{\rm{D}}\bra{\rm{D}}+\frac{1}{2}\ket{\rm{Sim}}\bra{\rm{Sim}}\right),\\
    \ket{w}&=\frac{1}{\sqrt{N}}\left(\ket{0\cdots01}+\ket{0\cdots10}+\cdots+\ket{1\cdots00} \right),
\end{align}
where
$\ket{w}$ represents a superposition state (W state) over $N$ modes. The state $\tilde{\rho}_t^{(\rm{nonlocal})}$ after time evolution under the given Hamiltonian for time $t$ is expressed as follows:

\begin{align}
    \tilde{\rho}_t^{(\rm{nonlocal})}=e^{-i\hat{H}t}\tilde{\rho}_0^{(\rm{nonlocal})}e^{i\hat{H}t}.
\end{align}
Expanding $\tilde{\rho}_t^{(\rm{nonlocal})}$ , it can be expressed as:
\begin{align}
     \tilde{\rho}_t^{(\rm{nonlocal})}&=\frac{1}{2}\ket{w_{{\rm{Sim}}}}\bra{w_{{\rm{Sim}}}}\otimes\ket{\rm{Sim}}\bra{\rm{Sim}}\notag \\
     &+\frac{1}{2}\ket{w_{\rm{D}}}\bra{w_{\rm{D}}}\otimes\ket{\rm{D}}\bra{\rm{D}},
\end{align}
where
we define $\ket{w_{{\rm{Sim}}}}$ and $\ket{w_{\rm{D}}}$ as follows:
\begin{align}
    &\ket{w_{{\rm{Sim}}}} = \frac{1}{\sqrt{N}}(\exp[-it\omega_N^{\prime}]\ket{0\cdots01}\\&+\exp[-it\omega_{N-1}^{\prime}]\ket{0\cdots10}  
    +\cdots+\exp[-it\omega_1^{\prime}]\ket{1\cdots00}),\\
    \notag&\ket{w_{\rm{D}}}
   =\frac{1}{\sqrt{N}}(\exp[-it\omega_N]\ket{0\cdots01}\\&+\exp[-it\omega_{N-1}]\ket{0\cdots10}  
    +\cdots+\exp[-it\omega_1]\ket{1\cdots00}).
\end{align}
The states after time evolution, $\tilde{\rho}_{{\rm{Sim}}}^{(\rm{nonlocal})}$ for $\ket{\rm{Sim}}$ and $\tilde{\rho}_{\rm{D}}^{(\rm{nonlocal})}$ for $\ket{{\rm{D}}}$, are given by:
\begin{align}
    \tilde{\rho}_{{\rm{Sim}}}^{(\rm{nonlocal})}&=\ket{w_{{\rm{Sim}}}}\bra{w_{{\rm{Sim}}}}\otimes\ket{\rm{Sim}}\bra{\rm{Sim}},\\
    \tilde{\rho}_{\rm{D}}^{(\rm{nonlocal})}&=\ket{w_{\rm{D}}}\bra{w_{\rm{D}}}\otimes\ket{\rm{D}}\bra{\rm{D}}.
\end{align}
The corresponding projection operators are defined as follows:
\begin{align}
    \hat{\tilde{P}}_{{\rm{Sim}}}^{(\rm{nonlocal})}&=\ket{w}\bra{w} \otimes \hat{1}_{N+1}, \\
    \hat{\tilde{P}}_{\rm{D}}^{(\rm{nonlocal})}&= (\hat{1}_{1,2,\cdots,N}-\ket{w}\bra{w}) \otimes \hat{1}_{N+1} .
\end{align}
If the system is projected by $\hat{\tilde{P}}_{{\rm{Sim}}}^{(\rm{nonlocal})}$, the phase plate is identified as $|\rm{Sim}\rangle $ . Conversely, 
we consider that,
if the system is projected onto $\hat{\tilde{P}}_{\rm{D}}^{(\rm{nonlocal})}$ , the phase plate is identified as $|{\rm{D}}\rangle$  .
The error probability $P_{\rm{err}}$ is adopted in the same form as in the local state case, as given by Equation \eqref{eq:ayateiFD}.
Similarly, the prior probabilities are assumed to be $p_{{\rm{Sim}}}=p_{\rm{D}}=1/2$.
We can calculate
$\tilde{P}_{\rm{err}}^{(\rm{nonlocal})}$
as follows:
\begin{align}
    \tilde{P}_{\rm{err}}^{(\rm{nonlocal})}=\frac{1}{2}\left( 1-\left|\braket{w|w_{{\rm{Sim}}}}\right|^2+\left|\braket{w|w_{\rm{D}}}\right|^2 \right).  \label{eq:hVShisiayahikyoku}
\end{align}
We can
calculate the values of $\left|\braket{w|w_{\rm{D}}}\right|^2$ and $\left|\braket{w|w_{{\rm{Sim}}}}\right|^2$
as follows:
\begin{align}
     \left|\braket{w|w_{\rm{D}}}\right|^2&=\left|\frac{1}{N}\sum_{j=1}^{N}\exp[it\omega_j]\right|^2,\\
     \left|\braket{w|w_{{\rm{Sim}}}}\right|^2&=\left|\frac{1}{N}\sum_{j=1}^{N}\exp[it\omega_j^{\prime}]\right|^2.
\end{align}

Let us derive
the statistical average of $\left|\braket{w|w_{\rm{D}}}\right|^2$ 
using the above equations.
\begin{align}
    \notag &\langle \left|\braket{w|w_{\rm{D}}}\right|^2 \rangle\\
    \notag&=\left\langle \left| \frac{1}{N}\sum_{j=1}^N \exp[i\hat{\theta}_j] \right|^2 \right\rangle \\
    \notag&=\left\langle \frac{1}{N^2}\sum_{j,j^{\prime}=1}^N \exp[i\hat{\theta}_j]\exp[-i\hat{\theta}_{j^{\prime}}] \right\rangle \\
    \notag&=\left\langle \frac{1}{N^2}\left( \sum_{j=j^{\prime},j=1}^N 1 + \sum_{j\neq j^{\prime}}  \exp[i\hat{\theta}_j]\exp[-i\hat{\theta}_{j^{\prime}}] \right)\right\rangle\\
    &=\frac{1}{N},
\end{align}
where 
$\hat{\theta}_j$ represents a classical random variable, and the phases chosen for different plates are assumed to be independent. Given that 
$\theta_j$ is
uniformly distributed between $0$ and $2\pi$, the statistical average of $\exp[i\hat{\theta}_j]\exp[-i\hat{\theta}_{j^{\prime}}]$ becomes 0.
Consequently, the statistical average value of $\left|\braket{w|w_{\rm{D}}}\right|^2$ becomes $\frac{1}{N}$.
Also,
the statistical average of $\left|\braket{w|w_{{\rm{Sim}}}}\right|^2$ is calculated.
\begin{align}
    \notag \langle \left|\braket{w|w_{{\rm{Sim}}}}\right|^2 \rangle=
    \notag&\left\langle \left| \frac{1}{N}\sum_{j=1}^N \exp[i\hat{\theta}_j^{\prime}] \right|^2 \right\rangle \\
    \notag&\simeq \left\langle \frac{1}{N^2} \left| \sum^{N}_{j=1}\left(1-i\hat{\theta}_j^{\prime}-\frac{1}{2}\hat{\theta}_j^{\prime2}\right) \right|^2 \right\rangle \\
    \notag &\simeq \left\langle  1-\frac{1}{N}\sum^N_{j=1}\hat{\theta}_j^{\prime2}+\frac{1}{N^2}\sum^N_{j=1}\hat{\theta}_j^{\prime2}  \right\rangle \\
    &= 1-\frac{1}{3}\frac{\pi^2}{M^2}+\frac{1}{3}\frac{1}{N}\frac{\pi^2}{M^2}.
\end{align}
Here, $\theta_1^{\prime},\theta_2^{\prime}\cdots\theta_N^{\prime}$ are assumed to be uniformly distributed within $[-\frac{\pi}{M},\frac{\pi}{M}]$, and $M$ is assumed to be sufficiently large. 
Using these results, the statistical average of $\tilde{P}_{\rm{err}}^{(\rm{nonlocal})}$ is obtained as:
\begin{align}
    \langle \tilde{P}_{\rm{err}}^{(\rm{nonlocal})} \rangle\simeq\frac{1}{2}\left( \frac{1}{3}\frac{\pi^2}{M^2}-\frac{1}{3}\frac{1}{N}\frac{\pi^2}{M^2}+\frac{1}{N} \right). \label{eq:hVShisiayahikyokukai}
\end{align}
Therefore, in the limit of sufficiently large $N$ , the value of $\tilde{P}_{\rm{err}}^{(\rm{nonlocal})}$ asymptotically approaches 0. This indicates that increasing the number of phase plates makes it possible to identify the properties of the phase plates 
almost without errors.

\subsubsection{Comparison}

\begin{figure}[h!t]
    \centering
    \includegraphics[width=0.5\textwidth]{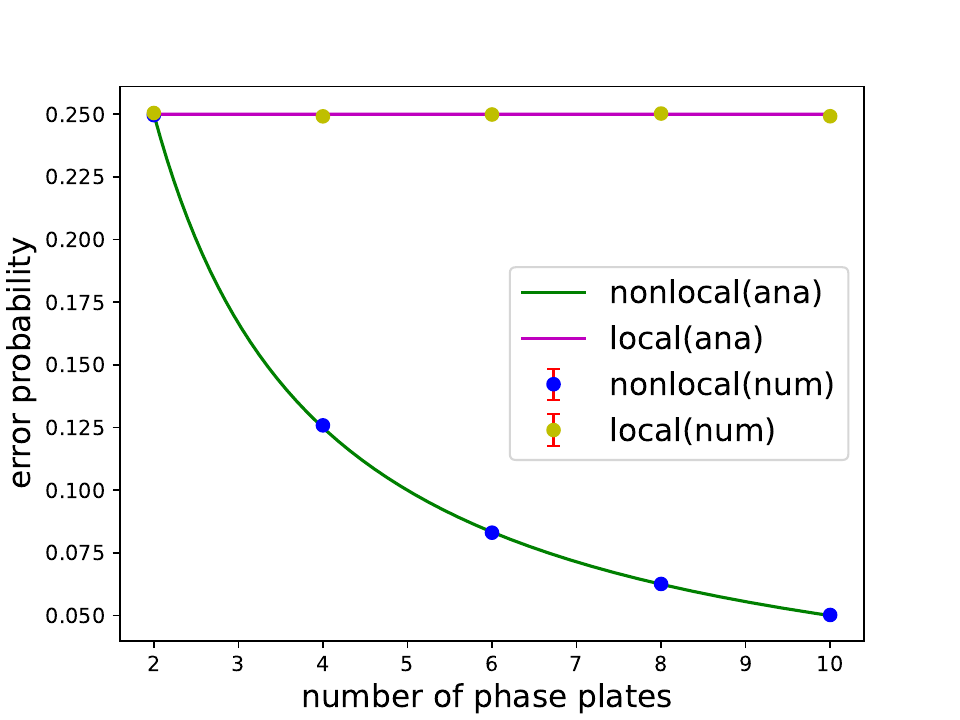}
    \caption{The plot shows the error probabilities for both the local and nonlocal states. The number of phase plates considered ranges from 2 to 10, with \( t = 1 \), the number of trials set to 100,000, and \( M = 10,000 \).  
    The numerical results obtained by repeatedly computing equation \eqref{eq:hVShisiayakyoku} and averaging the outcomes closely match the analytical results given by equation \eqref{eq:hVShisiayakyokuana} for the local state.  
    The numerical results obtained by repeatedly computing equation \eqref{eq:hVShisiayahikyoku} and averaging the outcomes closely match the analytical results given by equation \eqref{eq:hVShisiayahikyokukai} for the nonlocal state.  
    The standard error is included as error bars in the plot. However, the values are so small that the markers almost hide them.
    }
        \label{fig:FP_l_FD}
\end{figure}

\begin{figure}[h!t]
    \centering
    \includegraphics[width=0.5\textwidth]{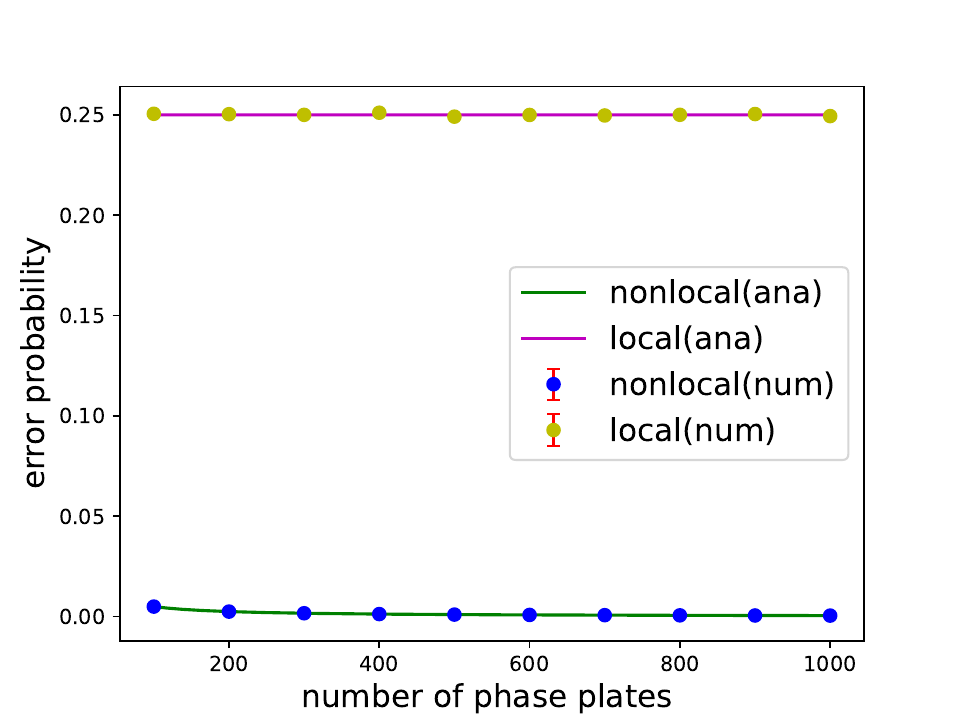}
    \caption{The plot shows the error probabilities for both the local and nonlocal states. The number of phase plates considered ranges from 100 to 1000, with \( t = 1 \), the number of trials set to 100,000, and \( M = 10,000 \).  
    The numerical results obtained by repeatedly computing equation \eqref{eq:hVShisiayakyoku} and averaging the outcomes closely match the analytical results given by equation \eqref{eq:hVShisiayakyokuana} for the local state.  
    The numerical results obtained by repeatedly computing equation \eqref{eq:hVShisiayahikyoku} and averaging the outcomes closely match the analytical results given by equation \eqref{eq:hVShisiayahikyokukai} for the nonlocal state.  
    The standard error is included as error bars in the plot. However, the values are so small that the markers almost hide them.
    }
    \label{fig:FP_m_FD}
\end{figure}

We compare the use of single photons in nonlocal states versus local states when interacting with phase plates. Fig.\ref{fig:FP_l_FD} and Fig.\ref{fig:FP_m_FD} illustrate the relationship between the error probability and the number of phase plates, $N$, for both cases. In Fig.\ref{fig:FP_l_FD}, the number of phase plates ranges from 2 to 10. The graph plots the analytical solution for the error probability in nonlocal states (Equation \eqref{eq:hVShisiayahikyokukai}), the numerically calculated error probability for nonlocal states (Equation \eqref{eq:hVShisiayahikyoku}), the analytical solution for the error probability in local states (Equation \eqref{eq:hVShisiayakyokuana}), and the numerically calculated error probability for local states (Equation \eqref{eq:hVShisiayakyoku}). The value of $M$ is set to 10,000. When there are two phase plates, no significant difference in error probability is observed. However, as the number of phase plates increases, the error probability for nonlocal states becomes lower than that for local states.

Fig.\ref{fig:FP_m_FD} extends this analysis to cases where the number of phase plates ranges from 100 to 1,000. The graph plots the error probability for nonlocal states (Equation \eqref{eq:hVShisiayahikyoku}) and includes the analytical solution derived under the assumption of sufficiently large $M$ (Equation \eqref{eq:hVShisiayahikyokukai}). For local states, the numerically calculated error probability (Equation \eqref{eq:hVShisiayakyoku})) is also plotted, along with its analytical solution derived under the assumption of sufficiently large $M$ (Equation \eqref{eq:hVShisiayakyokuana}). As the number of phase plates increases, the error probability for nonlocal states decreases and converges to nearly 0. Furthermore, the analytical solutions and numerical results show high agreement, demonstrating consistency of our analysis. 

\section{Conclusion}

In conclusion, we proposed a method to identify the properties of the phase plates by considering an
interaction of a single photon with $N$ phase plates, which is considered as one of a quantum sensor network (QSN). 
Specifically, we considerd two cases. One of them is distinguishing phase plates that impart the same phase from those that impart random phases in the range $[0,2\pi]$.
The other is
distinguishing phase plates that impart the
random phases in the range $[0,2\pi]$ from those that impart random phases in the range $[-\delta, \delta]$ where $\delta$ is a small constant.
These are considered as a binary classification.
We compared the identification accuracy when using a nonlocal single-photon state 
with that when using
a local single-photon state.
We demonstrated that the use of a nonlocal single-photon state leads to improved identification accuracy.
These findings highlight new possibilities for QSN and contribute to the broader application of quantum information processing with photons. Moreover, such binary classification is relevant to algorithms designed for 
quantum computing, such as the Deutsch-Jozsa algorithm ~\cite{deutsch1992rapid}, as well as variational quantum machine learning algorithms for intermediate-scale quantum computing ~\cite{cerezo2021variational,endo2021hybrid,bharti2022noisy,wang2024comprehensive}, which have been actively studied ~\cite{mitarai2018quantum}. Since our approach involves binary classification using QSN, it has the potential to bridge interdisciplinary research between QSN and quantum computing, further underscoring its significance.

This work is supported by JSPS KAKENHI (Grant Number 23H04390), JST Moonshot (Grant Number JPMJMS226C), CREST (JPMJCR23I5), and Presto JST (JPMJPR245B).
YT is partially supported by JST [Moonshot R\&D -- MILLENNIA Program] Grant Number JPMJMS2061.

\bibliography{ref}

\end{document}